\documentclass[onecolumn,secnumarabic,amssymb, nobibnotes, aps, prd]{revtex4}
\usepackage{amsmath} \usepackage{amssymb}
\usepackage{graphicx} 
\usepackage{epstopdf}
\usepackage{amsmath}
\usepackage{amsmath,amssymb,amsthm,amsfonts,mathrsfs,bm,verbatim}
\usepackage{graphicx,subfigure}

\newcommand{\bea}{\begin{eqnarray}}
\newcommand{\eea}{\end{eqnarray}}

\begin{document}

\title{Thermodynamic geometric analysis of BTZ black hole  in  $f(R, \phi)$  Gravity }%

\author{Wen-Xiang Chen$^{a}$}
\affiliation{Department of Astronomy, School of Physics and Materials Science, GuangZhou University, Guangzhou 510006, China}
\author{Yao-Guang Zheng}
\email{hesoyam12456@163.com}
\affiliation{Department of Astronomy, School of Physics and Materials Science, GuangZhou University, Guangzhou 510006, China}

\begin{abstract}
In this work, we obtain the thermodynamic geometry of the extended black hole solution under $f(R)$ gravity. This is the foundation of our work. We also obtain the exact form of the f(R) model for some solutions. In the thermodynamic analysis, we calculate the thermodynamic quantities such as temperature and entropy of these solutions and compare them with the corresponding BTZ black holes. After that, we study the Ruppeiner geometry, such as the f(R)-gravity (extended black hole solution) of the BTZ black hole, and analyze the thermodynamic comparison of the Ruppeiner geometry with the van der Waals equation for its phase transition process.

 \text { KEYwords: } $f(R, \phi)$ -gravity; Ruppeiner geometry;BTZ black hole

\end{abstract}

\maketitle

\section{Introduction}
In the $(2+1)$-dimension, one of the first exact black hole solutions of general relativity (GR) is the Ba\~{n}ados, Teitelboim, and Zanelli (BTZ) black holes. It is known that in the $(2+1)$ dimension, the Weyl tensor vanishes by definition, and in the absence of matter we have $R_{\mu\nu}=0$, so we cannot have a black hole solution. To avoid this problem, a negative cosmological constant was introduced into the theory, allowing black hole solutions to exist in spacetime with a negative constant curvature.

After the discovery of the BTZ black hole, several $(2+1) $dimensional black hole solutions coupled with scalar fields have been obtained. Black holes with ubiquitous regular scalar fields have been studied. Then, more $(2+1)$-dimensional hairy black hole solutions are discussed. Conformally coupled scalar and Abelian gauge fields are found in asymptotically AdS$(2+1)$-dimensional black hole solutions. $(2+1)$ dimensional charged black holes and scalar hairs are derived, where the scalar potential is not fixed but derived from Einstein's equations. Exact $(2+1)$-dimensional black holes with non-minimum coupled scalar fields are discussed, where arbitrary coupling constants break conformal invariance. The $(2+1)$-dimensional dilaton gravitational sum studies the modification of the dilaton/scalar to the BTZ black hole. Recently, a regular black hole solution with a real scalar field coupled to a Maxwell field was constructed by dual transformation, and a $(2+1)$-dimensional regular black hole with a nonlinear electrodynamic source was studied\cite{1,2,3,4,5,6,7,8,9,10,11,12,13,14}.

Ruppeiner geometry is based on fluctuations. When the microstructure of the thermodynamic system under study is unknown, Ruppeiner geometry provides us with a powerful tool for exploring the microstructure of the thermodynamic system. The metric of Ruppeiner geometry is\cite{15,16,17}
\begin{equation}
ds^{2}=-\frac{\partial^{2} S}{\partial X^{\alpha} \partial X^{\beta}} \Delta X^{\alpha} \Delta X^{\ beta}
\end{equation}
Where $S$ is the entropy of the thermodynamic system, $\Delta X^{\alpha}=X^{\alpha}-X_{0}^{\alpha}$ is the thermodynamic quantity $X^{\alpha}$ deviates from equilibrium The fluctuation of the thermodynamic quantity $X_{0}^{\alpha}$ at the time. The physical explanation of this metric is very clear: the farther away from the equilibrium state in phase space, the smaller the probability of being at that point. For ordinary thermodynamic systems, if the thermodynamic system studied is Fermion, the curvature scalar of Ruppeiner geometry is positive; if the thermodynamic system studied is Boson, the curvature scalar of Ruppeiner geometry is negative; for ideal classical gases, the curvature of Ruppeiner geometry is The scalar is zero. Further research shows that if the curvature scalar is positive, the microscopic interaction of the system is repulsive; if the curvature scalar is negative, the microscopic interaction of the system is attractive.

The specific form of the Van der Waals equation is:
\begin{equation}
\left(p+\frac{a^{\prime}}{v^{2}}\right)\left(v-b^{\prime}\right)=k T
\end{equation}
in the formula
- $p$ is the pressure of the gas
- $a^{\prime}$ is a phenomenological parameter for measuring intermolecular gravity
- $b^{\prime}$ is the volume contained in a single molecule itself
- $v$ is the average space occupied by each molecule (i.e. the volume of the gas divided by the total number of molecules);
- $k^{k}$ is a Perez ordinary number
- $T$ is absolute temperature

This paper also studies the thermodynamics and Ruppeiner geometry of the BTZ black hole -$f(R, \phi)$ gravity\cite{18}. The Ruppeiner geometry of the angular momentum-fixed ensemble is curved, while the Ruppeiner geometry of the pressure-fixed ensemble is flat. This paper reviews the interpretation of Ruperna geometry, but has no conclusive results. The Ruperner geometry results for BTZ black hole-f(R) gravity further support that the curvature scalar of Ruperner geometry does sometimes encode information about black hole stability and we analyze the thermodynamic comparison of the Ruppeiner geometry with the van der Waals equation for its phase transition process.

\section{Description of the system(CONFORMAL BTZ BLACK HOLE)}
We first briefly review conformally dressed black holes. The role of the theory consists of a Ritchie scalar, a negative cosmological constant, and a conformally coupled scalar field, namely\cite{8,9}.
\begin{equation}
S=\frac{1}{2} \int d^{3} x \sqrt{-g}\left\{\frac{R+2 l^{-2}}{\kappa}-\partial_{\mu} \phi \partial^{\mu} \phi-\frac{1}{8} R \phi^{2}\right\}
\end{equation}
where we will use $\kappa=8 \pi G=1$ for simplicity throughout the paper. By variation one can obtain the Einstein equation.

We assume the metric ansatz with $g_{t t} g_{r r}=-1$
\begin{equation}
d s^{2}=-b(r) d t^{2}+b(r)^{-1} d r^{2}+r^{2} d \theta^{2}
\end{equation}
where $b(r)$ is the only degree of freedom and it can be obtained from the Ricci scalar
\begin{equation}
b(r)=\frac{r^{2}}{l^{2}}-\frac{c_{1}}{r}+c_{2}
\end{equation}
while from the $t t$ and $r r$ components of the Einstein equation we can get the scalar field
\begin{equation}
\phi(r)=\frac{1}{\sqrt{c_{3} r+c_{4}}} .
\end{equation}
Substituting them into the $\theta \theta$ component of the Einstein equation
\begin{equation}
b(r)=\frac{r^{2}}{l^{2}}+\frac{B^{2}(-2 B-3 r)}{l^{2} r},
\phi(r)=\sqrt{\frac{8 B}{r+B}},
\end{equation}
where $B=c_{4} / c_{3}>0$

Since Newton's gravitational constant $\mathrm(G)$ is related to the dimension of space-time, Chapter 3 considers four-dimensional space-time, adopts the geometric unit system, and takes $\mathrm(G)=1$. But it is more convenient for three-dimensional space-time, generally take $8\mathrm{G}=1$. In this chapter, all $\mathrm{G}$ are written out. In order to distinguish the following Gibbs functions, this chapter records Newton's gravitational constant G as $G_{N}$.

\section{Thermodynamic parameters of BTZ black holes in $f(R, \phi)$ theory}
We can immediately integrate the Ricci scalar $f_{R}(r)$ to obtain the general form of $f(R, \phi)$  theory\cite{9}
\begin{equation}
f_{R}(r)=1+\alpha r \rightarrow f(R)=R+\alpha \int^{R} r(R) d R+C,
\end{equation}
Where $C$ is an integral constant, the unit is $[L]^{-2}$, which is related to the cosmological constant. This expression shows that, in addition to the Einstein-Hilbert term, a geometric correction term appears, while there is no immediate scalar field in the $f(R)$ model.
Then we can solve the metric function as
\begin{equation}
b(r)=-\frac{3 B^{2}}{l^{2}(\alpha B+1)^{2}}-\frac{2 B^{3}}{l^{2 } r(\alpha B+1)}+\frac{6 \alpha B^{2} r}{l^{2}(\alpha B+1)^{3}}+r^{2}\left (\frac{1}{l^{2}}+\frac{6 \alpha^{2} B^{2}}{l^{2}(\alpha B+1)^{4}} \ln \left(\frac{r}{\alpha l(B+r)+l}\right)\right)
\end{equation}
where $l$ is the AdS radius displayed as an integration constant.
The first law of thermodynamics is
\begin{equation}
d M=T d S+V d P+\Omega d J
\end{equation}
where the pressure
\begin{equation}
P=\frac{1}{4 \pi} \frac{1}{l^{2}}
\end{equation}
thermodynamic volume
\begin{equation}
V=\frac{\partial M}{\partial P}=\frac{2 G_{N}}{\pi} S^{2}
\end{equation}

Hawking temperature can be calculated as
\begin{equation}
T_{H}=\frac{b^{\prime}(r_{h})}{4 \pi}=\frac{3 B^{2}(B+r_{h})}{2 \pi l^{2} r_{h}^{2}(\alpha B+\alpha r_{h}+1)}
\end{equation}
which uses the relation $b\left(r_{h}\right)=0$.

We can calculate the entropy of the black hole under the gravitational force of f(R), whose non-minimum coupling is
\begin{equation}
S=-\left.\frac{1}{4} \int d \theta \sqrt{r_{h}^{2}}\left(\frac{\partial \mathcal{L}}{\partial R_{ \alpha \beta \gamma \delta}}\right)\right|_{r=r_{h}} \hat{\varepsilon}_{\alpha \beta} \hat{\varepsilon}_{\gamma \delta }
\end{equation}
where $\hat{\varepsilon}_{\alpha \beta}$ is the binormal to the horizontal plane $[80], \mathcal{L}$ is the Lagrangian of the theory, and
\begin{equation}
\frac{\partial \mathcal{L}}{\partial R_{\alpha \beta \gamma \delta}}|_{r=r_{h}}=\frac{1}{2} (\frac{f_{R}(r_{h})}{2}-\frac{1}{16} \phi(r_{h})^{2})(g^{\alpha \gamma} g^{\beta \delta}-g^{\beta \gamma} g^{\alpha \delta})
\end{equation}
Finally, we can get the entropy formula of our theory
\begin{equation}
S=\pi r_{h}\left(\frac{f_{R}\left(r_{h}\right)}{2}-\frac{1}{16} \phi\left(r_{h} \right)^{2}\right)=\frac{\mathcal{A}}{4} f_{R_{\mathrm{total}}}\left(r_{h}\right)
\end{equation}
Replacing $f_{R_{\text {total }}}$ with an explicit expression, we have
\begin{equation}
S=\frac{1}{2} \pi r_{h}\left(1+\alpha r_{h}-\frac{B}{B+r_{h}}\right)
\end{equation}
In fact, $r_{h}$ here also changes with the choice of $B, l$ and $\alpha$.

\begin{equation}
d s^{2}=-\frac{\partial^{2} S(r_{h},B)}{\partial X^{\alpha} \partial X^{\beta}} \Delta X^{\alpha} \Delta X^{\beta}
\end{equation}
\begin{equation}
g_{ij}=
\begin{pmatrix}
  \pi\left(a+\frac{B^{2}}{(B+r h)^{3}}\right)&  -\frac{B \pi r h}{(B+r h)^{3}} \\
-\frac{B \pi r h}{(B+r h)^{3}}    &  \frac{\pi r h^{2}}{(B+r h)^{3}}
\end{pmatrix}
\end{equation}
The curvature scalar of thermodynamic geometry is
\begin{equation}
R(S)=\frac{1}{2 a \pi r_h^{2}}
\end{equation}
We see the possibility of a phase transition in the BTZ black hole at that time. When a tends to 0, the solution exists and the curvature scalar diverges.When
\begin{equation}
d s^{2}=-\frac{\partial^{2} S(r_{h},a)}{\partial X^{\alpha} \partial X^{\beta}} \Delta X^{\alpha} \Delta X^{\beta}
\end{equation}
\begin{equation}
R(S)=0.
\end{equation}

We find the functional formula of P-V:
\begin{equation}
P = \frac{\operatorname{T_H}\left(a r_h^{2}-V\right)^{2} V}{6 r_h^{2}\left(r_h+a r_h^{2}-V\right)^{2}}
\end{equation}

\begin{figure}[htp]
\includegraphics[width=14cm,height=7cm]{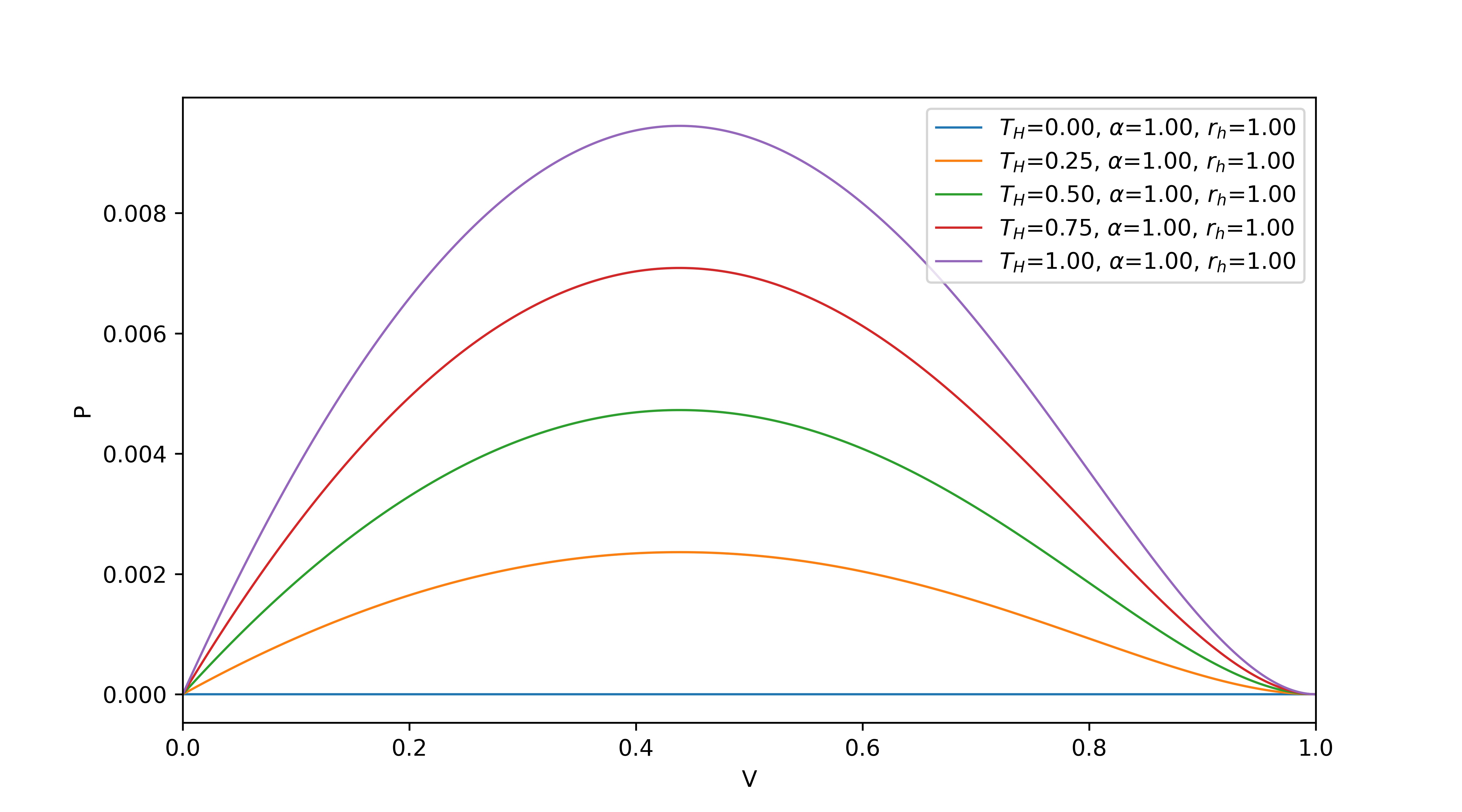}
\caption{ $1$}
\end{figure}
\begin{figure}[htp]
\includegraphics[width=14cm,height=7cm]{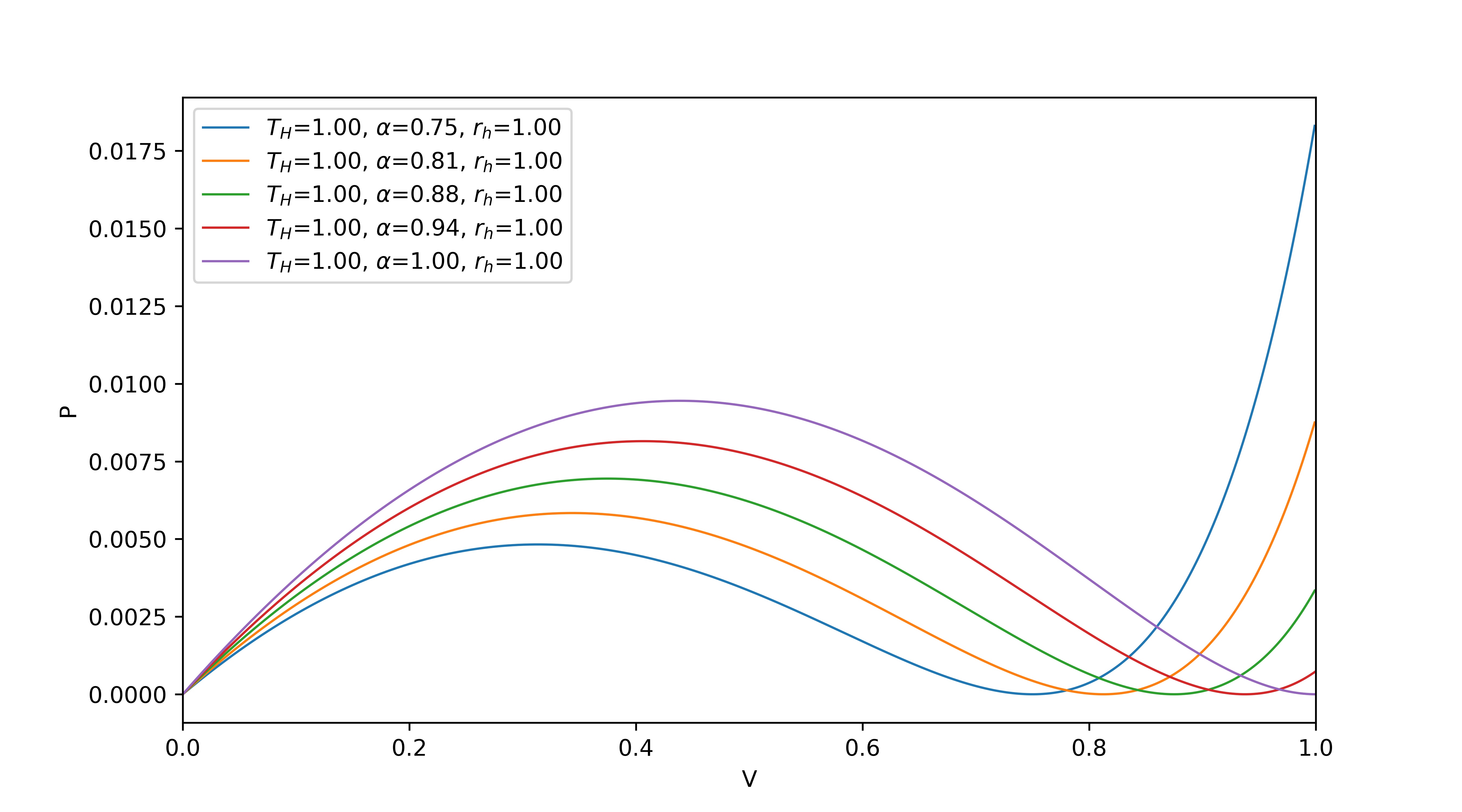}
\caption{ $ 2$}
\end{figure}
\begin{figure}[htp]
\includegraphics[width=14cm,height=7cm]{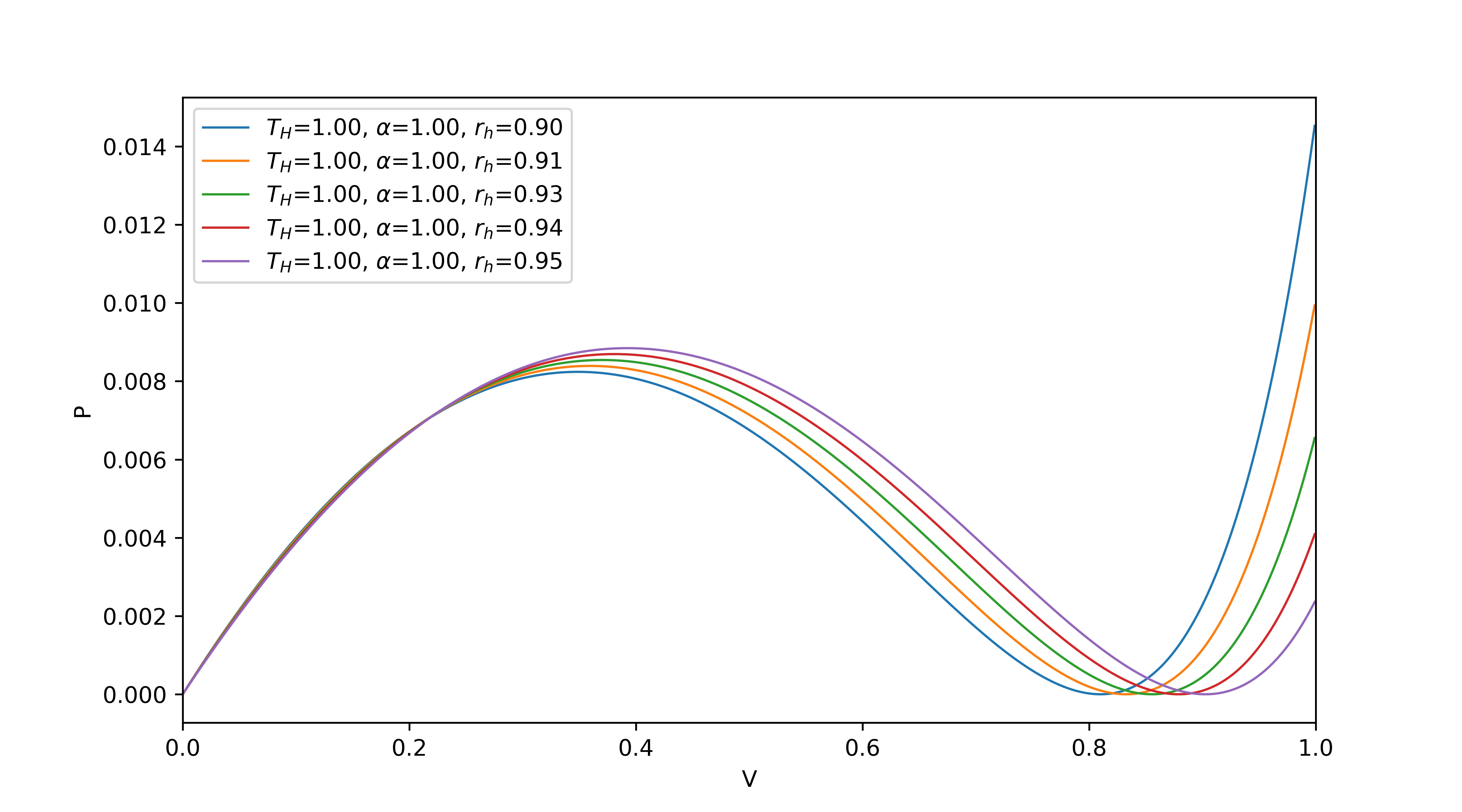}
\caption{ $ 3$}
\end{figure}
\begin{figure}[htp]
\includegraphics[width=14cm,height=7cm]{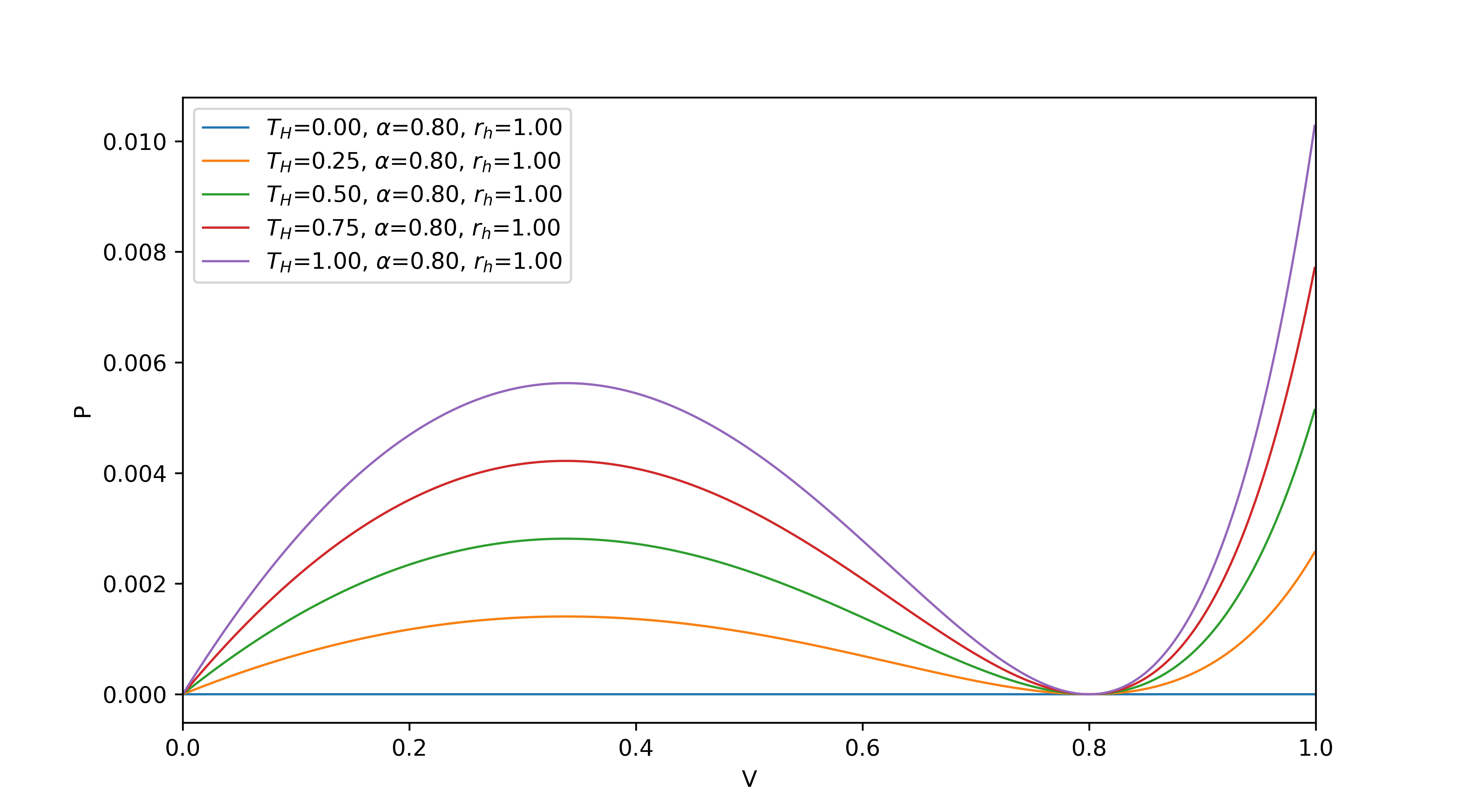}
\caption{ $4$}
\end{figure}
\begin{figure}[htp]
\includegraphics[width=14cm,height=7cm]{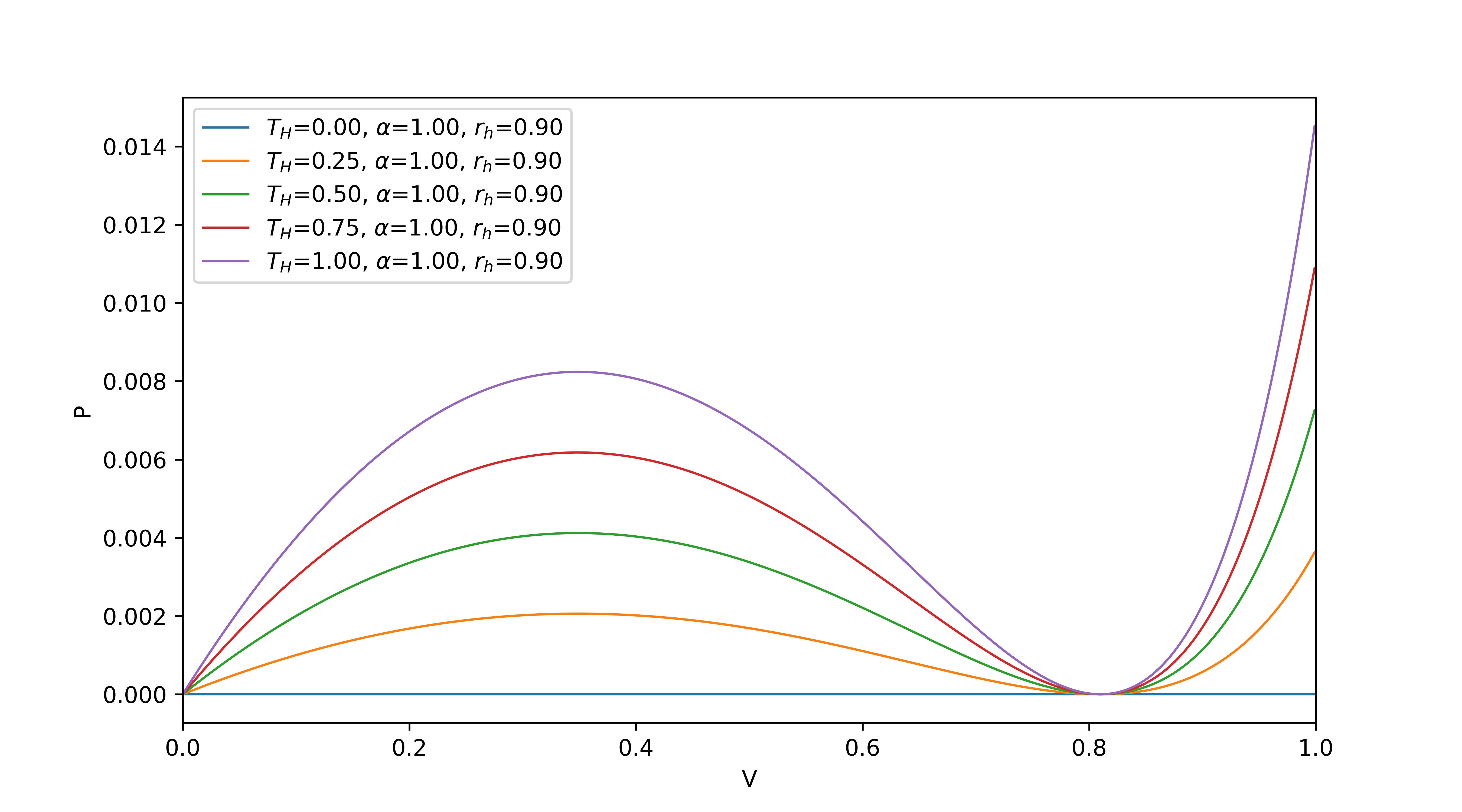}
\caption{ $ 5$}
\end{figure}
We found that when the temperature is very small, the phase transition phenomenon occurs, with similar results as when analyzed with R geometry.

We try to come up with an idea. If B and $a$ are regarded as a constant, then the P-V relationship can be regarded as the $P-r_h$ relationship:(when B=1,$a$=1,FIG.6)
\begin{equation}
P = \frac{r_h^{2}(2+r_h) \mathrm{T_H}}{6(1+r_h)}
\end{equation}
\begin{figure}[htp]
\includegraphics[width=14cm,height=7cm]{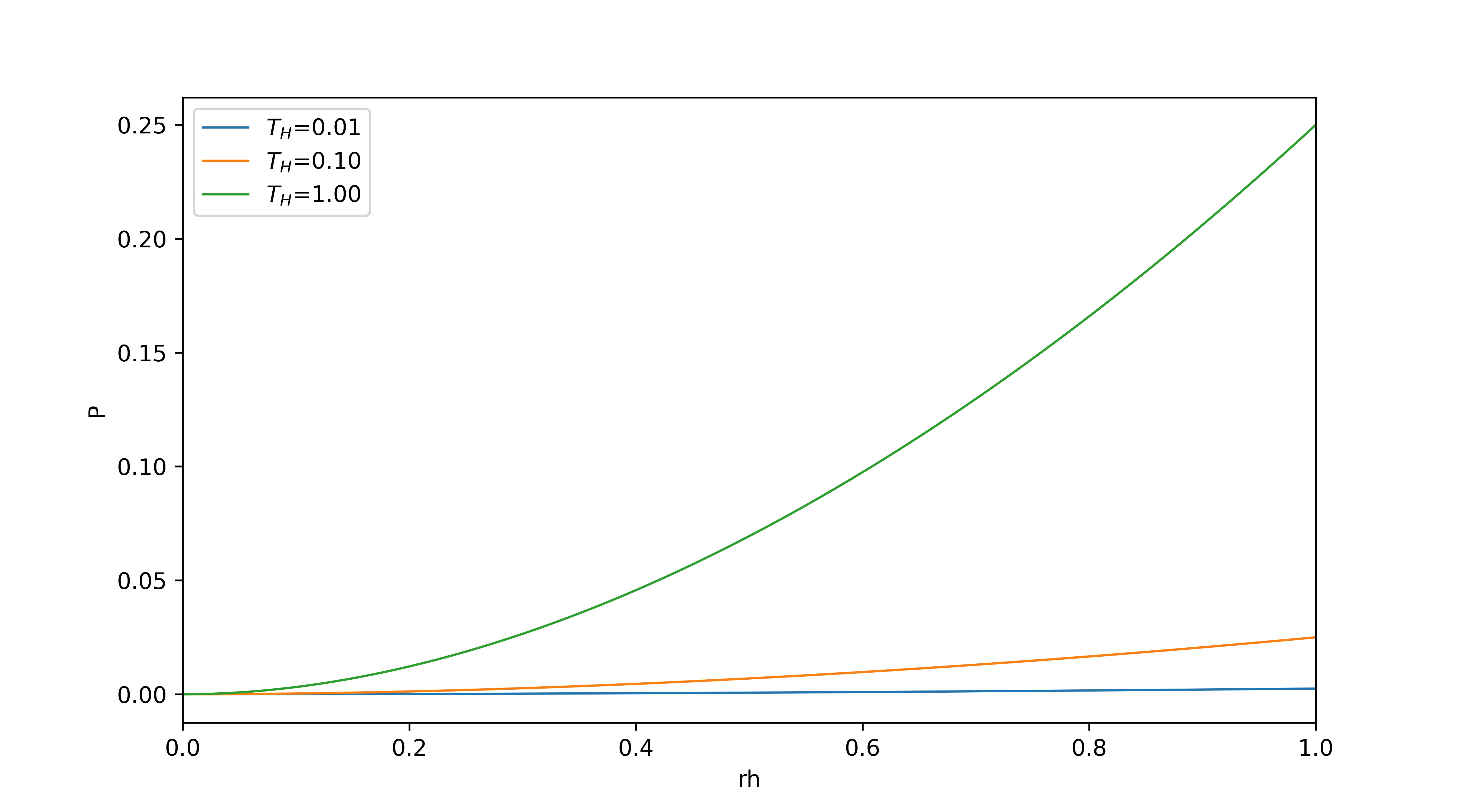}
\caption{ $ 6$}
\end{figure}
We see that there is no evidence that there is a phase transition in the BTZ black hole at this time, so the BTZ black hole at that time can be regarded as an ideal gas.

Compare the case of a general BTZ black hole\cite{5,6,7}:

We list the general thermodynamic geometry analysis of the ordinary BTZ black hole, we see that the curvature scalar of its thermodynamic geometry is always greater than 0, which means that there is no phase transition. From this, perhaps we know from analysis that when under f(R) gravity (including its extended solution), the BTZ black hole may have a phase transition.
\begin{equation}
\begin{aligned}
d s_{R}^{2} &=\frac{1}{T} d s_{W}^{2}=\frac{1}{T} \frac{\partial^{2} M}{\partial X^{\alpha} \partial X^{\beta}} d X^{\alpha} d X^{\beta} \\
&=\frac{1}{T}\left[\frac{\partial^{2} M}{\partial S^{2}} d S^{2}+\frac{\partial^{2} M}{\partial S \partial P} d S d P+\frac{\partial^{2} M}{\partial P^{2}} d P^{2}\right] .
\end{aligned}
\end{equation}we get that,
\begin{equation}
\begin{array}{ll}
g_{S S}=\frac{1}{T} \frac{\partial^{2} M}{\partial S^{2}}=\frac{1}{T} \frac{\partial T}{\partial S}, & g_{S P}=g_{P S}=\frac{1}{T} \frac{\partial^{2} M}{\partial S \partial P}=\frac{1}{T} \frac{\partial T}{\partial P} \\
g_{P P}=0, & g=\operatorname{det}\left(g_{\mu \nu}\right)=-g_{S P}^{2}
\end{array}
\end{equation}
\begin{equation}
R(M)=\frac{\pi^{2}}{16 G_{N}^{3}} \frac{J^{2}}{T S^{4}}
\end{equation}The curvature scalar is always greater than zero. Therefore, the microscopic interaction is repulsion.

\section{Summary and Discussion}
This paper also studies the thermodynamics and Ruppeiner geometry of the BTZ black hole -$f(R, \phi)$ gravity. The Ruppeiner geometry of the angular momentum-fixed ensemble is curved, while the Ruppeiner geometry of the pressure-fixed ensemble is flat. This paper reviews the interpretation of Ruperna geometry, but has no conclusive results. The Ruperner geometry results for BTZ black hole-f(R) gravity further support that the curvature scalar of Ruperner geometry does sometimes encode information about black hole stability and we analyze the thermodynamic comparison of the Ruppeiner geometry with the van der Waals equation for its phase transition process.We list the general thermodynamic geometry analysis of an ordinary BTZ black hole, and we see that its thermodynamic geometry curvature scalar is always greater than 0, which means there is no phase transition. From this, perhaps we know from the analysis that under f(R) gravity (including its extended solution), the BTZ black hole may undergo a phase transition.

{\bf Acknowledgements:}\\
This work is partially supported by  National Natural Science Foundation of China(No. 11873025).

\end{document}